\documentclass[aps,prl,twocolumn,preprintnumbers, superscriptaddress,
               floatfix,nofootinbib]{revtex4}
\usepackage{graphicx}
\usepackage{epstopdf}
\usepackage{mathrsfs}
\usepackage{amssymb}
\usepackage{amsmath}
\usepackage{bm}
\usepackage{verbatim}
\usepackage{color}
\usepackage{multirow}

\def\stacksymbols #1#2#3#4{\def\theguybelow{#2}
    \def\vp{\lower#3pt}
    \def\sp{\baselineskip0pt\lineskip#4pt}
    \mathrel{\mathpalette\intermediary#1}}

\def\intermediary#1#2{\vp\vbox{\sp
     \everycr={}\tabskip0pt
     \halign{$\mathsurround0pt#1\hfil##\hfil$\crcr#2\crcr
              \theguybelow\crcr}}}

\def\be{\begin{equation}}
\def\ee{\end{equation}}
\def\bea{\begin{eqnarray}}
\def\eea{\end{eqnarray}}

\def\sp{\;\;\;,\;\;\;}

\def\lsim{\raise0.3ex\hbox{$\;<$\kern-0.75em\raise-1.1ex\hbox{$\sim\;$}}}
\def\gsim{\raise0.3ex\hbox{$\;>$\kern-0.75em\raise-1.1ex\hbox{$\sim\;$}}}

\def\inbar{\,\vrule height1.5ex width.4pt depth0pt}

\def\IC{\relax\hbox{$\inbar\kern-.3em{\rm C}$}}
\def\IQ{\relax\hbox{$\inbar\kern-.3em{\rm Q}$}}
\def\IR{\relax{\rm I\kern-.18em R}}
 \font\cmss=cmss10 \font\cmsss=cmss10 at 7pt
\def\IZ{\relax\ifmmode\mathchoice
 {\hbox{\cmss Z\kern-.4em Z}}{\hbox{\cmss Z\kern-.4em Z}}
 {\lower.9pt\hbox{\cmsss Z\kern-.4em Z}}
 {\lower1.2pt\hbox{\cmsss Z\kern-.4em Z}}\else{\cmss Z\kern-.4em Z}\fi}

\def\comment#1{}

\def\u1x{U(1)_X}
\newcommand{\nc}{\newcommand}
\nc{\LL}{L}
\nc{\vv}{\tilde{v}}
\nc{\ccdot}{\!\cdot\!}
\nc{\gsm}{G_{SM}}
\nc{\vfive}{\mathbf{5}\oplus\mathbf{\overline{5}}}
\nc{\vten}{\mathbf{10}\oplus\mathbf{\overline{10}}}
\nc{\zhol}{Z^{\rm hol}}
\nc{\xfb}{\,{\rm fb}}

\begin{document}

\preprint{LPT--Orsay--16--12}
\preprint{UMN--TH--3515/16}
\preprint{FTPI--MINN--16/05}

\vspace*{1mm}

\title{Vacuum Stability and Radiative Electroweak Symmetry Breaking \\
in an SO(10) Dark Matter Model}

\author{Yann Mambrini}
\email{yann.mambrini@th.u-psud.fr}
\affiliation{Laboratoire de Physique Th\'eorique 
Universit\'e Paris-Sud, F-91405 Orsay, France.
 }
\author{Natsumi Nagata}
\email{nagat006@umn.edu}
\affiliation{
 William I.~Fine Theoretical Physics Institute, 
       School of Physics and Astronomy,
            University of Minnesota, Minneapolis, MN 55455, USA}
\author{Keith A. Olive}
\email{olive@physics.umn.edu}
 \affiliation{
 William I.~Fine Theoretical Physics Institute, 
       School of Physics and Astronomy,
            University of Minnesota, Minneapolis, MN 55455, USA}
\author{Jiaming Zheng}
\email{zheng@physics.umn.edu}
 \affiliation{
 William I.~Fine Theoretical Physics Institute, 
       School of Physics and Astronomy,
            University of Minnesota, Minneapolis, MN 55455, USA}

\begin{abstract} 
 Vacuum stability in the Standard Model is problematic as the Higgs
 quartic self-coupling runs negative at a renormalization scale of about
 $10^{10}$ GeV. We consider a non-supersymmetric SO(10) grand
 unification model for which gauge coupling unification is made possible through an intermediate
scale gauge group, $G_{\rm int}=\text{SU}(3)_C\otimes
 \text{SU}(2)_L\otimes \text{SU}(2)_R \otimes \text{U}(1)_{B-L}$.
 $G_{\rm int}$ is broken by the vacuum expectation value of a {\bf 126}
 of SO(10) which not only provides for neutrino masses through the
 seesaw mechanism but also preserves a discrete $\mathbb{Z}_2$ that 
 can account for the stability of a dark matter
 candidate, here taken to be the Standard Model singlet component of a
 bosonic {\bf 16}. We show that in addition to these
 features the model insures the positivity of the Higgs quartic
 coupling through its interactions to the dark matter multiplet and  {\bf 126}. 
 We also show that the Higgs mass squared runs negative triggering
 electroweak symmetry breaking. Thus, the vacuum stability is achieved
 along with radiative electroweak symmetry breaking and captures two
 more important elements of supersymmetric models without low-energy
 supersymmetry. The conditions for perturbativity of quartic couplings
 and for radiative electroweak symmetry breaking lead to tight upper and
 lower limits on the dark matter mass, respectively, and this dark
 matter mass region (1.35--2 TeV) can be probed in future direct detection experiments. 

\end{abstract}

\maketitle


\maketitle


\setcounter{equation}{0}




{\em Introduction.}---With the discovery of the Higgs boson at both the ATLAS
\cite{HiggsATLAS} and CMS \cite{HiggsCMS} detectors, the Standard Model
(SM) of particle physics appears to be very well established. However, as
yet, there is no verified explanation for neutrino masses, and the
nature of dark matter (DM) remains elusive. Both signal the need for
beyond the SM physics. The experimental value of the Higgs mass, $m_h =
125.09\pm 0.24$~GeV \cite{125}, also points to new physics at some
higher energy scale. The Higgs quartic self-coupling, $\lambda$, runs
toward negative values at high energy. The lower limit on $m_h$ to
ensure the posititivity of $\lambda$ out to the Planck scale is $129.4
\pm 1.8$~GeV \cite{degrassi}, which appears to be violated. DM is often
introduced in supersymmetric extensions of the SM with $R$-parity
\cite{EHNOS}. In supersymmetric models, the problem of vacuum stability associated with the Higgs
quartic coupling is avoided as the tree level coupling is determined by
a combination of the gauge couplings and is positive definite. In
addition, supersymmetric models offer a mechanism for triggering
electroweak symmetry breaking via radiative effects \cite{ewsb}.

In the absence of supersymmetry, the instability in the Higgs potential occurs
at a renormalization scale of about $10^{10}$ GeV for $m_h \simeq 125$ GeV.
Therefore, one might anticipate new physics playing a role at this
intermediate scale or below \cite{EliasMiro:2011aa, OlegHiggs}, such as the
seesaw mechanism for generating neutrino masses, which has long been associated
with an intermediate scale \cite{seesaw}. This mechanism is very
naturally realized in SO(10) grand unified theories (GUTs)
\cite{Georgi:1974my,GN2} where the right-handed neutrino is
included as the SM singlet component of the {\bf 16} of SO(10) and
incorporates a full generation of matter fields in a single
representation. SO(10) contains several 
subgroups, $G_{\rm int}$, which contain the SM gauge group as a a subgroup,
and it is well known that the symmetry breaking scale, $M_{\rm int}$, of
$G_{\rm int}$ can be determined by requiring that the gauge couplings
unify at a single scale $M_{\rm GUT} > M_{\rm int}$  \cite{GN2,masiero,
moqz, mnoqz, noz}.  

Stable dark matter can also be incorporated in SO(10) models in a
straightforward way \cite{kkr,kkr2, moqz, mnoqz, noz, Evans:2015cqq}. As a
rank-five group, SO(10) includes an additional U(1) symmetry, which is
assumed to be broken at the intermediate scale. If the Higgs field that
breaks this additional U(1) symmetry belongs to a ${\bf 126}$-dimensional 
representation, then a discrete $\mathbb{Z}_2$ symmetry is preserved at
low energies \cite{Kibble:1982ae}.  If we restrict our attention to
relatively small representations ($ \le \bf{210}$), the ${\bf 126}$
Higgs field  leaving a $\mathbb{Z}_2$ symmetry is the only possibility
for a discrete symmetry \cite{DeMontigny:1993gy,mnoqz}. For example, a
scalar dark matter candidate will be stabilized by the $\mathbb{Z}_2$
symmetry  if it is a member of either a {\bf 16} or {\bf 144}
representation.

In this paper, we show that the two aforementioned attributes of
supersymmetric extensions to the SM, namely, vacuum stability and
radiative electroweak symmetry breaking, are also natural consequences
of SO(10) models with an intermediate scale gauge group. For
definiteness, we consider here a SM singlet dark matter candidate originating
from a single {\bf 16} of SO(10) as in model ${\tt SA}_{3221}$ in
Ref.~\cite{noz} based on the intermediate gauge group
$\text{SU}(3)_C\otimes \text{SU}(2)_L\otimes \text{SU}(2)_R \otimes
\text{U}(1)_{B-L}$. In this model, the intermediate scale is found to be
$M_{\rm int} \simeq 10^9~{\rm GeV}$ and is small enough to allow the
couplings of the ${\bf 126}$ Higgs field to the SM Higgs to lift the
Higgs quartic coupling through the threshold corrections before it turns
negative. The presence of the singlet scalar DM at low energies also
deflects the running of the Higgs quartic coupling. Moreover, we show
that the negative mass squared needed for electroweak symmetry breaking
runs positive due the coupling of the Higgs field with the DM singlet. 

The requirement for the radiative electroweak symmetry breaking imposes
a lower bound on the DM-Higgs coupling. This then leads to a lower
limit on the DM mass if one assumes that the thermal relic abundance of
the DM agrees with the observed DM density $\Omega_{\rm DM} h^2 \simeq
0.12$ \cite{Planck15}. On the other hand, perturbativity of the couplings in the model
gives an upper limit on the DM-Higgs coupling, and thus on the DM mass.
As a result, a finite DM mass region is
allowed by these two conditions. We find that this mass range
can be probed in the XENON1T experiment \cite{Aprile:2015uzo}.

{\em An exemplary SO(10) model with stable dark matter.}---When one combines the number of possible intermediate scale gauge groups
with the multitude of choices for dark matter and Higgs representations
in an SO(10) model, one may think that the amount of freedom one has for
model building is enormous. However, in practice, when one imposes the
conditions that i) gauge coupling unification occurs,
ii) that the intermediate scale is found to be below the GUT
scale, and iii) that the GUT scale is high enough so that the proton
lifetime exceeds current experimental bounds, only a handful of possible
models survive \cite{mnoqz,noz}. Furthermore, since any dark matter candidate must be
part of a larger SO(10) representation, that multiplet must be split,
putting further constraints on the possible choice of field content. 
 
In this paper, we choose one example of a scalar dark matter model
with an intermediate scale gauge group given by  $G_{\rm int} =
\text{SU}(3)_C\otimes \text{SU}(2)_L\otimes \text{SU}(2)_R \otimes
\text{U}(1)_{B-L}$. We will examine the model labeled ${\tt SA}_{3221}$
in Ref.~\cite{noz} for which the dark matter is a scalar singlet originating
in a {\bf 16} of SO(10). In addition to SM fields, the model employs a
{\bf 45} (or {\bf 210}) to break SO(10) to $G_{\rm int}$ when the ({\bf
15, 1, 1}) component (under $\text{SU(4)}_C \otimes \text{SU(2)}_L
\otimes \text{SU(2)}_R$) acquires a vacuum expectation value (vev). The
intermediate scale gauge group is subsequently broken when the color
singlet, right-handed triplet sitting in the {\bf 126} acquires a
vev. All other components of the {\bf 126} are expected to have GUT
scale masses. In addition to an explicit (GUT scale) mass term for the
{\bf 16}, the scalar multiplet can have mass contributions from its
couplings to the Higgs {\bf 45} and {\bf 126}. An explicit calculation
of the fine-tuning needed to obtain a TeV scale mass for the singlet
scalar dark matter candidate can be found in Appendix C of
Ref.~\cite{noz}. In the example given there, all members of the {\bf 16} are
GUT scale except the scalar analog of $e_R$ (${\tilde e_R}$), which has
an intermediate scale mass, and ${\tilde \nu_R}$, which has a weak scale
mass.

\begin{figure}[ht!]
\begin{center}
\includegraphics[height=70mm]{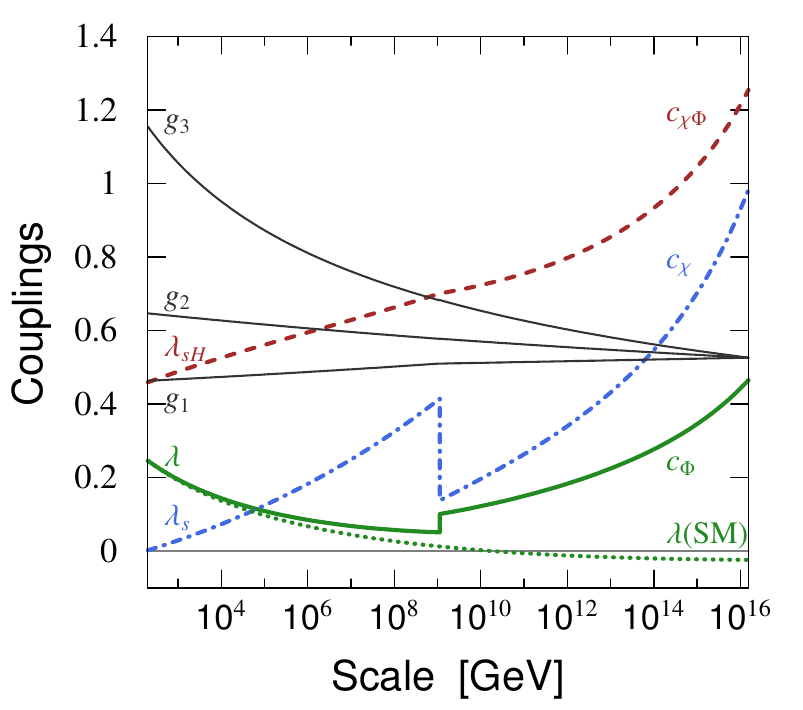}
\end{center}
\caption{\it Running of the quartic couplings of the Higgs field, for
 selected inputs. The green solid, brown dashed, and blue dash-dotted
 lines show the running of $\lambda$, $\lambda_{sH}$, and $\lambda_s$,
 respectively, while the green dotted curve shows the running of
 $\lambda$ in the SM. The gauge coupling running is also shown in thin
 black lines. Above the intermediate scale, the running of $c_\Phi,
 c_\chi$, and $c_{\chi\Phi}$ is shown using the matching conditions in
 \eqref{match}. The free parameters are chosen as
 follows: at $Q=m_t$, $\lambda_s=0$, and $\lambda_{sH}=0.46$ (which
 corresponds to $m_{\rm DM} \simeq 1.5$~TeV); at $M_{int}$,
 $\tilde{c}_\Phi = c^\prime_\Delta =c_{\Phi\Delta} = c_{\chi\Delta} =
 c^\prime_{\chi\Delta} = c^\prime_{\chi\Phi} =0$, and $c_\Delta
 = -c^\prime_{\Phi \Delta}  = -m_{\chi\Delta}/v_R =
 0.05$. The non-zero couplings are taken so that the low-energy mass
 spectrum we consider here is realized.  
}  
\label{fig:compare}
\end{figure}

{\em Renormalization group evolution of the Higgs couplings and
 masses.}---The renormalization group evolution between the weak scale and
intermediate scale is almost identical to the SM. The only difference
comes from the inclusion of the SM singlet dark matter candidate, $s
 \equiv {\rm Re}[{\tilde \nu_R}]$. 
Below the intermediate scale, the scalar potential is relatively simple,   
 \begin{equation}
V_{\rm blw}= \mu^2 |H|^2  +\frac{1}{2}\mu^2_s s^2 +
 \frac{\lambda}{2}|H|^4 +\frac{\lambda_{sH}}{2}|H|^2 s^2 +
 \frac{\lambda_s}{4!}s^4 ~.
\end{equation}
In many ways, this resembles the minimal dark matter model often
referred to as the Higgs portal \cite{hp, Davoudiasl:2004be}.
The mass of our dark matter candidate is given by $m_{\rm DM}^2 =
\lambda_{sH} v^2/2 + \mu_s^2$. Furthermore, fixing the dark matter mass
will also fix $\lambda_{sH}$ at the weak scale (taken here to be $m_t$)
through the relic density (assuming standard thermal freeze-out):
$m_{\rm DM} \simeq 3.3 \lambda_{sH}~{\rm TeV}$. In this paper, we
compute the DM relic density using {\tt micrOMEGAS}
\cite{Belanger:2014vza}. The evolution of
the Higgs quartic coupling in the SM with and without the inclusion of
the scalar $s$ is shown in Fig.~\ref{fig:compare}
by the green solid and dotted curves, respectively. The renormalization
group equations (RGEs) are run at the two-loop level\footnote{We use the
three-loop RGEs for the top Yukawa and Higgs quartic couplings. We also
include the two-loop electroweak threshold corrections according to
Ref.~\cite{degrassi}. We use the $\overline{\rm MS}$
scheme up to the intermediate scale and switch to the $\overline{\rm
DR}$ scheme at $M_{\rm int}$. } and one sees that the SM quartic coupling
runs negative just above $10^{10}$ GeV \cite{degrassi} without the
scalar contribution. With the scalar contribution, the running of $\lambda$ would
remain positive out to the GUT scale.
Note that at the intermediate scale (determined by the conditions for 
gauge coupling unification; the running of the gauge couplings in ${\tt
SA}_{3221}$ is shown by thin black lines in Fig.~\ref{fig:compare})
$M_{\rm int} \simeq 10^9$ GeV, $\lambda > 0$. Gauge coupling unification
also determines the GUT scale to be $M_{\rm GUT}\simeq 1.5\times
10^{16}$~GeV, which is high enough to evade the proton decay limit. Also
shown is the running of $\lambda_s$ (blue dash-dotted line) and $\lambda_{sH}$
(brown dashed line). 

Above the intermediate scale, it is necessary to include in addition to $s$
the right-handed doublet $\chi({\bf 1},{\bf 1},{\bf 2},1)$ which
contains $s$, the Higgs triplet $\Delta({\bf 1},{\bf 1},{\bf 3},2)$
residing in the {\bf 126}, two heavy complex fields in addition to the
SM Higgs doublet which all sit in a complex $\Phi({\bf 1},{\bf 2},{\bf
2},0)$, and finally the three right-handed neutrinos sitting in the
fermionic {\bf 16} matter representations. Above the intermediate scale,
we write $\Phi= (\phi_1, \tilde{\phi}_2)$, 
$\tilde{\Phi} \equiv \sigma_2 \Phi^* \sigma_2$ ($\sigma_a$ are the Pauli
matrices), $\chi = (\chi^+, \chi^0)^T$, and 
\begin{equation}
\Delta=
\begin{pmatrix}
  \Delta^+/\sqrt{2} &  \Delta^{++} \\
\Delta^0               &  -\Delta^+/\sqrt{2}
\end{pmatrix}
~,
\end{equation}
where $\phi_i=(\phi^0_i, \phi^-_i)^T$ is an ${\rm SU}(2)_L$ doublet;
$\tilde{\phi} \equiv i\sigma_2\phi^*$. Then, a quartic potential can be
written as 
\begin{align}
V_{\rm abv}^{(4)}&=
\frac{c_\Delta}{2} \left({\rm\bf tr}(\Delta^\dagger\Delta)\right)^2
+\frac{c'_\Delta}{4} {\rm\bf tr}\left(\Delta\Delta\right){\rm\bf tr}
\left(\Delta^\dagger\Delta^\dagger\right)  \nonumber\\
&+\frac{c_\Phi}{2}\big({\rm\bf tr}(\Phi^\dagger\Phi)\big)^2
+\frac{\tilde{c}_\Phi}{4}{\rm\bf tr}(\tilde{\Phi}^\dagger\Phi){\rm\bf tr}(\Phi^\dagger\tilde{\Phi}) \nonumber\\
& +c_{\Phi\Delta}{\rm\bf tr}(\Delta^\dagger\Delta){\rm\bf tr}(\Phi^\dagger\Phi)
+\frac{c_\chi}{2}|\chi|^4 
+ c_{\chi\Phi} |\chi|^2 {\rm\bf tr}(\Phi^\dagger\Phi)
\nonumber\\
&+ c_{\chi\Delta} |\chi|^2 {\rm\bf tr}(\Delta^\dagger\Delta)
+ c^\prime_{\chi\Delta}\chi^\dagger[\Delta^\dagger,\Delta]\chi
 \nonumber\\
& +c'_{\Phi\Delta}{\rm\bf tr}\left( \Phi^\dagger\Phi [\Delta^\dagger,\Delta]\right)   +c'_{\chi\Phi}\chi^\dagger\Phi^\dagger\Phi\chi +  \dots ~.
\end{align} 
Note that we have only included those quartic couplings which can be
generated through RGE evolution, with the exception of the last two; 
$c^\prime_{\Phi\Delta}$  is needed to split the masses of the
two-Higgs doublet, $\Phi$, while $c^\prime_{\chi\Phi}$ is
induced by the $c^\prime_{\Phi\Delta}$ term via RGE effects.

The quartic terms that contain two powers of $\Delta$, as well as the
cubic coupling (see Eq.~\eqref{eq:quadcub}), produce non-trivial tree-level threshold corrections at $M_{\rm int}$, after $\Delta$ acquires a vev and the heavy fields are integrated out,
\begin{align}
\lambda &=
 c_\Phi-\frac{(c_{\Phi\Delta}+c'_{\Phi\Delta})^2}{c_\Delta}~, 
 \nonumber\\
\lambda_{sH} &= c_{\chi\Phi} 
-\frac{(c_{\Phi\Delta}+c'_{\Phi\Delta})[m_{\chi\Delta}+(c_{\chi\Delta}-c_{\chi\Delta}^\prime)v_R]}{
 c_\Delta v_R} ~,
\nonumber\\
\lambda_s &= 3c_\chi-3
\frac{[m_{\chi\Delta}+v_R(c_{\chi\Delta}-c'_{\chi\Delta})]^2}{c_\Delta
 v_R^2} ~,
\label{match}
\end{align}
where $\langle \Delta \rangle = v_R T_-$ with $T_- \equiv (\sigma_1 -i
\sigma_2)/2$. 
As is well known, these threshold effects always go in the direction of
benefiting vacuum stability \cite{EliasMiro:2011aa}. The evolution of
the quartic couplings, $c_\Phi, c_\chi$, and $c_{\chi\Phi}$ above the
intermediate scale are also shown in Fig.~\ref{fig:compare} using the
matching conditions in  \eqref{match}. We use the one-loop
RGEs for these quartic couplings. Although we do not explicitly display
the running of all quartic terms above the intermediate scale, we have
checked that, although some run negative (notably $c^\prime_\Delta$),
the couplings satisfy sufficient conditions which guarantee stability of
the vacuum up to the GUT scale. 

The quadratic and cubic parts (which can lead to mass terms) of the potential can be written as 
\begin{align}
V_{\rm abv}^{(2,3)}&=m^2_\chi |\chi|^2+m_\Phi^2{\rm\bf tr}(\Phi^\dagger\Phi)
+m_\Delta^2{\rm\bf tr}(\Delta^\dagger\Delta) \nonumber \\
&+m_{\chi\Delta}\left( \tilde{\chi}^\dagger \Delta^\dagger \chi \right)
+{\rm h.c.}
 ~,
\label{eq:quadcub}
\end{align}
where we take $m_{\chi \Delta}$ to be real for simplicity. 
The relevant matching conditions with the weak scale mass parameters are
\begin{align}
\mu_s^2&=m_\chi^2 + \left( c_{\chi\Delta} -
 c'_{\chi\Delta} \right)v_R^2 +2m_{\chi\Delta}v_R~,
\nonumber\\
\mu^2&=m_\Phi^2+\left( c_{\Phi\Delta}+c'_{\Phi\Delta}\right)v_R^2 ~,
\end{align}
where the low-energy fields are related to the high-energy fields as
$\phi_1=H$ and $\chi^0=(s+ i a)/\sqrt{2}$. 

The running of $\lambda_s$ receives a large contribution from $\lambda_{sH}$,
$d\lambda_s/d\ln Q  = 12 \lambda_{sH}^2/(4\pi)^2 + \cdots$, and thus by
demanding perturbativity of the couplings $(\lambda_i \lesssim 1/\beta_i$, where $\beta_i$ is
a relevant beta-function coefficient) up to the
intermediate scale, we can set an upper bound on $\lambda_{sH}
\lesssim1.3$. However, requiring perturbativity of the $c_i$'s above the intermediate scale 
places a stronger bound on $\lambda_{s}(M_{\rm int}) \lesssim 2.4$ which 
requires $\lambda_{sH}(m_t) \lesssim 0.9$. Non-zero values for other couplings 
further push the upper limit to $\lambda_{sH}(m_t) \lesssim 0.6$ in order to avoid singularities in the 
RGEs. 
Since $\lambda_{sH}$ controls the annihilation cross section
for $s$, $\sigma_{\rm ann} v_{\rm rel} \simeq \lambda_{sH}^2/16 \pi
m_{\rm DM}^2$, and the relic density is proportional to $1/\langle
\sigma_{\rm ann} v_{\rm rel} \rangle$, the upper limit on $\lambda_{sH}$
corresponds to an upper limit to the DM mass $m_{\rm DM} \lesssim 2$~TeV, similar to that in the 
minimal dark matter model \cite{Davoudiasl:2004be} without an intermediate scale.

The Higgs mass parameter, $\mu^2$, must be negative in order to break
the electroweak symmetry, and in the SM, $\mu^2$ remains negative as it
is run up to high energies. The presence of the dark matter scalar,
however, affects the running as $d\mu^2/d\ln Q =  \lambda_{sH}
\mu_s^2/(4\pi)^2 + \cdots$ and causes $\mu^2$ to run positive at higher
renormalization scales \cite{kkr}. In other words, the dark matter candidate can
induce radiative electroweak symmetry similar to the mechanism in
supersymmetric models \cite{ewsb}. As the running of $\mu$ depends on
the combination $\lambda_{sH} \mu_s^2$, we can obtain a minimum value for
$\mu_s$ (and hence $m_{\rm DM}$), which is independent of the relic
density constraint, by maximizing $\lambda_{sH}$. We find that for
$\lambda_{sH} = 0.6$, $\mu^2 > 0$ at the intermediate scale (at 1~TeV)
when $\mu_s \gtrsim 360$~GeV ($1150$~GeV), corresponding to
$m_{\rm DM} \gtrsim 380$~GeV (1160~GeV). Here, we set $\lambda_s
(m_t) = 0$. Taking the limits on
$\lambda_{sH}$ from the perturbativity of $\lambda_s$ and the limit on
$\mu_s$ from the requirement of radiative electroweak symmetry breaking,
we find that the dark matter mass must lie in a restricted range (when
demanding the more natural choice of symmetry breaking at 1 TeV) $m_{\rm DM} = 1.2$--2~TeV.

\begin{figure}[ht!]
\begin{center}
\includegraphics[height=70mm]{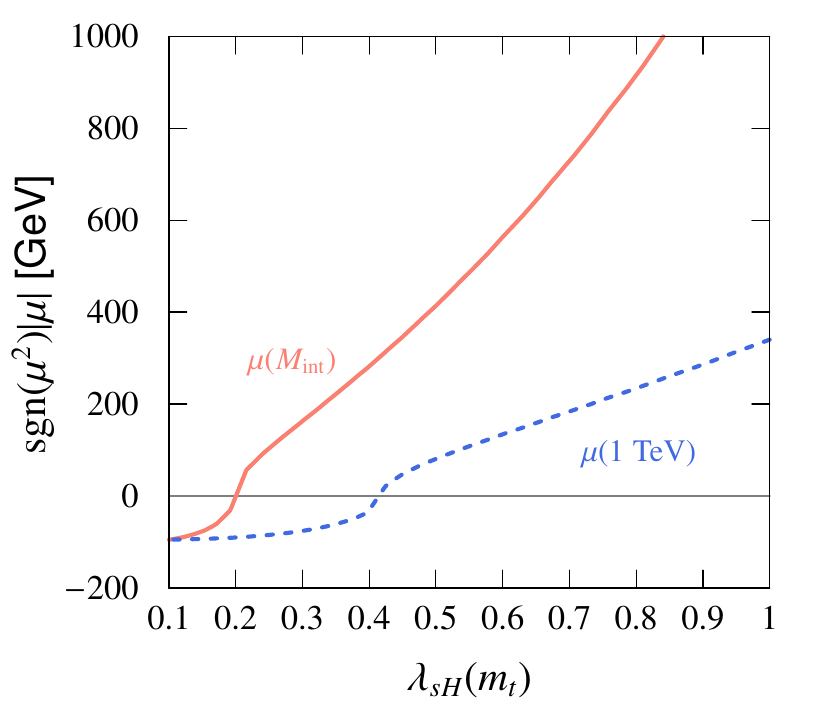}
\end{center}
\caption{\it The value of ${\rm sgn}(\mu^2)|\mu|$ for $Q = M_{\rm int}$
  and $1$~TeV as a function of $\lambda_{sH}(m_t)$. $m_{\rm DM}$ at the weak
  scale is determined from the requirement for the thermal relic
  abundance using $m_{\rm DM} \approx 3.3 \lambda_{sH}$ TeV. 
}  
\label{fig:mass}
\end{figure}

When one imposes the constraint from the relic density, we obtain somewhat stronger
bounds on $\lambda_{sH}$. In Fig.~\ref{fig:mass}, we show the value of
${\rm sgn}(\mu^2)|\mu|$ for $Q = M_{\rm int}$ and $1$~TeV as a function
of $\lambda_{sH}(m_t)$. Here, again, we set $\lambda_s(m_t) = 0$.
As one can see, when $Q = M_{\rm int}$, we
have $\lambda_{sH} (m_t) > 0.2$ corresponding to $m_{\rm DM} > 670$~TeV
and when $Q = 1$ TeV, we have $\lambda_{sH} (m_t) > 0.41$ corresponding
to $m_{\rm DM} > 1.35$~TeV. 

\begin{figure}[ht!]
\begin{center}
\includegraphics[height=70mm]{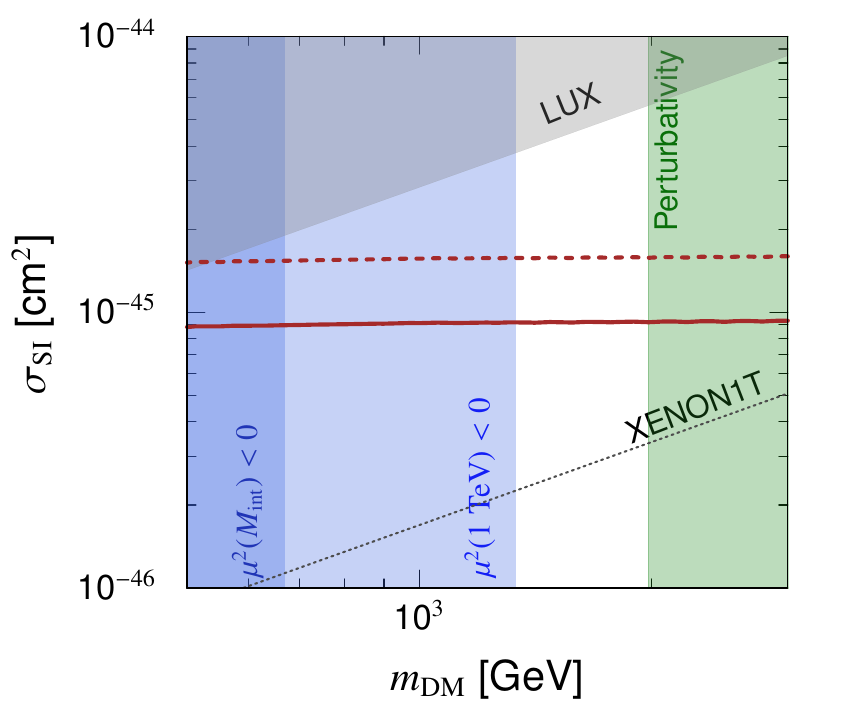}
\end{center}
\caption{\it The SI DM-nucleon scattering cross section as a function of
 $m_{\rm DM}$. Here, $\lambda_{sH}$ is determined from the relic density
 condition. 
}  
\label{fig:sicross}
\end{figure}

The singlet DM candidate in our model can be probed in DM direct detection
experiments. In Fig.~\ref{fig:sicross}, we show the spin-independent
(SI) DM-nucleon scattering cross section $\sigma_{\rm SI}$ as a function
of the DM mass. Here, we require the relic density condition to
determine $\lambda_{sH}$. The lower solid (upper dashed) brown line
shows the result for which we use the nucleon matrix elements given in
Ref.~\cite{Abdel-Rehim:2016won} (Ref.~\cite{Ellis:2008hf}). In either
case, we obtain $\sigma_{\rm SI} \simeq 10^{-45}~{\rm cm}^2$. The gray
shaded region is excluded by the current limit from the LUX experiment
\cite{Akerib:2013tjd}. We also show the projected sensitivity of
XENON1T \cite{Aprile:2015uzo} by the black dotted line. We find that all
of the DM mass range can be probed at this experiment.

{\em Summary.}---We have presented an SO(10) model with gauge coupling unification
made possible through an intermediate scale at $\simeq 10^9$ GeV. SO(10) is broken
to $G_{\rm int}=\text{SU}(3)_C\otimes
 \text{SU}(2)_L\otimes \text{SU}(2)_R \otimes \text{U}(1)_{B-L}$ when the  right-handed triplet
 in the {\bf 126} obtains a vev. In this model, the lightest member of a complex scalar {\bf 16} is
 stable and plays the role of our dark matter candidate, $s$. 
The specific example discussed here can be viewed as a UV completion of the minimal (scalar) 
dark matter model. 
We have shown that, in addition to gauge coupling unification and a dark matter candidate,
unlike the case in the SM, vacuum stability is achieved up to the GUT scale, and
radiative electroweak symmetry breaking is triggered by the interactions of the dark 
matter and the SM Higgs. The latter result taken together with the requirement
of perturbative couplings to the GUT scale limits the DM mass to lie between
1.35--2 TeV. This mass range should be probed in future direct detection experiments.

\noindent {\bf Acknowledgments. } 
This  work was supported by the Spanish MICINN's
Consolider-Ingenio 2010 Programme  under grant  Multi-Dark {\bf CSD2009-00064}, and 
the LIA-TCAP of CNRS
and the France-US PICS no. 06482.
 Y.M.  acknowledges partial support from the European Union FP7 ITN INVISIBLES (Marie
Curie Actions, PITN- GA-2011- 289442) and  the ERC advanced grants  
 Higgs@LHC. The work of N. N. and K.A.O. was supported in part
by DOE grant DE--SC0011842 at the University of Minnesota.

\end{document}